\documentstyle[prl,aps,multicol,amssymb,amsmath,epsf,psfrag]{revtex}

\begin{document}

\title{Practical quantum repeaters with linear optics and
  double-photon guns}

\author{Pieter Kok\cite{pieter}, Colin P.\ Williams, and Jonathan P.\ Dowling}
\address{Quantum Computing Technologies Group, Section 367 \\
  Jet Propulsion Laboratory, California Institute of Technology \\
  Mail Stop 126-347, 4800 Oak Grove Drive, Pasadena, California 91109} 

\maketitle

\begin{abstract}
 We show how to create practical, efficient, quantum repeaters,
 employing double-photon guns, for long-distance optical quantum
 communication. The guns create polarization-entangled photon pairs on
 demand. One such source might be a semiconducter quantum dot, which
 has the distinct advantage over parametric down-conversion that the
 probability of creating a photon pair is close to one, while the
 probability of creating multiple pairs vanishes. The swapping and
 purifying components are implemented by polarizing beam splitters and
 probabilistic optical CNOT gates. 

 \medskip

\end{abstract}

\medskip

\begin{multicols}{2}


One year ago Knill, Laflamme, and Milburn demonstrated that efficient
quantum computing is possible, in principle, with {\em only} linear
optics and projective measurements \cite{knill01}. In particular, when
arbitrarily many auxiliary modes are available, one does not need the very
weak nonlinearities assumed to be essential for these purposes. 
Subsequently, it was shown that the same resources (linear optics and
projective measurements) can be used to create highly nonclassical
number states \cite{lee02,fiurasek01}. These states are important for
applications such as quantum lithography and quantum interferometry
\cite{boto00,dowling98}. In addition, we showed that projective
measurements enable interferometric quantum non-demolition
measurements, again with only linear optics \cite{kok02}. In this
Letter, we continue this research program and show how one can make a
practical quantum repeater with this technique. 

Quantum repeaters are essential for single-photon optical quantum
comunication over distances longer than the attenuation length of the
channels used \cite{briegel98,dur99}. Repeaters employ a combination
of entanglement swapping \cite{bennett93} and entanglement
purification or distillation \cite{bennett96}; i.e., multiple pairs of
degraded entangled states are condensed into (fewer) maximally
entangled states, after which swapping is used to extend the (now
maximal) entanglement over greater distances. Both entanglement
distillation and swapping have been demonstrated experimentally
\cite{kwiat01,pan01}. In this Letter, we present a practical protocol
for optical quantum repeaters based on linear optics and a
double-photon gun.  


Until now, the source for polarization-entangled photon pairs has mostly
consisted of parametric down-converters, where a strong pump laser
is sent through a nonlinear crystal. The interaction between the laser
and the crystal results in entangled photon pairs. However, the output
of these devices are not clean, maximally entangled, two-photon
states, but rather a coherent superposition of multiple pairs. 

Suppose the effective interaction Hamiltonian of a parametric
down-converter is given by $H = i\kappa \hat{L}_+ - i\kappa^*
\hat{L}_-$, where $\hat{L}_+ = \hat{a}^{\dagger}_H \hat{b}^{\dagger}_V
- \hat{a}^{\dagger}_V \hat{b}^{\dagger}_H = \hat{L}_-^{\dagger}$
\cite{kok00}. Here, $\hat{a}^{\dagger}$ and $\hat{b}^{\dagger}$ are
the usual creation operators of the two optical modes and $H$ and $V$
are orthogonal polarization directions. The operator, $\hat{L}_+$
($\hat{L}_-$) is the creation (annihilation) operator for entangled
photon pairs (in this case polarization singlets). The outgoing state
of a spontaneous parametric down-converter is then given by
$|\Psi_{\rm out}\rangle = \exp(iHt/\hbar) |0\rangle =
\sum_{n=0}^{\infty} {\mathcal{N}}_n (\epsilon L_+)^n |0\rangle$. This
expression is obtained by normal ordering $\exp(iHt/\hbar)$, where
$\epsilon\equiv \hat{\boldsymbol{\kappa}}\tanh\kappa$ and
${\mathcal{N}}_n^{-1}$ is the multiple-pair normalization
$\sqrt{n!(n+1)!}$, which is analgous to the normalization factor
$\sqrt{n!}$ of the ordinary creation operator $\hat{a}^{\dagger}$. To
complete the analogy, the operators $\hat{L}_{\pm}$ satisfy the
commutation relations associated with the $su(1,1)$ algebra:
$[\hat{L}_-,\hat{L}_+] = 2 \hat{L}_0$ and $[\hat{L}_0,\hat{L}_{\pm}] =
\pm \hat{L}_{\pm}$, where $2\hat{L}_0 \equiv \hat{a}^\dagger_H
\hat{a}_H + \hat{a}^\dagger_V \hat{a}_V + \hat{b}^\dagger_H \hat{b}_H
+ \hat{b}^\dagger_V \hat{b}_V + 2$. The complex number $\epsilon$ is
the probability amplitude of creating  the maximally entangled
state. Therefore, down-converters only produce single pairs when
$|\epsilon|\ll 1$, and by far the major contribution to the state is
the vacuum $|0\rangle$. 

For large-scale applications, such as a quantum repeater, there is a
more serious drawback to down-conversion. One typically needs many
entangled photon pairs, which would require, say, $N$ down-converters to
fire in unison. This happens with probability $|\epsilon|^{2\, N}$. 
However, with approximately the same probability the first
down-converter produces $N$ photon pairs, while the others produce
nothing. Worse still, {\em any} distribution of $N$ photon pairs
scales proportional to $|\epsilon|^{2\, N}$. As shown in Refs.\
\cite{kok00} and \cite{braunstein98}, this may seriously affect the
performance of most applications.

We would therefore like to have a source with the following
properties: (1) whenever we push the button of our entanglement
source, we produce, on demand, a polarization-entangled photon-pair;
(2) the fidelity of the output of our entanglement source must be very
close to one. A source with these properties we call a double-photon
gun.

\begin{figure}[h]
  \begin{center}
  \begin{psfrags}
     \psfrag{Bragg mirrors}{Bragg mirrors}
     \psfrag{p-type}{p-type}
     \psfrag{n-type}{n-type}
     \psfrag{qdot}{qdot}
     \psfrag{photons}{photons}
     \epsfxsize=8in
     \epsfbox[-50 20 880 170]{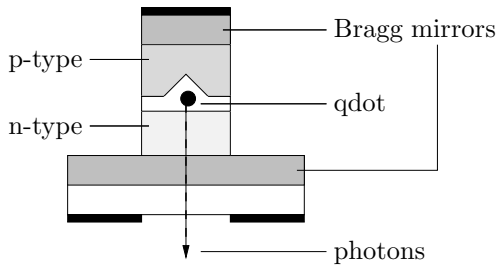}
  \end{psfrags}
  \end{center}
  \caption{The entanglement source. A quantum dot separating p-type
     and n-type GaAs is sandwiched between two Bragg
     mirrors. Electron-hole recombination will result in the creation
     of an entangled photon pair. Due to the Pauli exclusion principle,
     multiple pair production is suppressed. The efficiency of this
     proposed source is predicted to reach values up to 90\%.} 
  \label{fig1}
\end{figure}

One entanglement source that very nearly meets our requirements has
been proposed by Yamamoto and co-workers (see Fig.\ \ref{fig1})
\cite{benson00}. A quantum dot separating p-type and n-type GaAs is
sandwiched between two Bragg mirrors. The entire structure is
therefore an optical microcavity, and electron-hole recombination will
result inthe creation of an entangled photon pair. Critically, due to
Pauli's exclusion principle, only one electron and one hole are
recombined at a time, resulting in at most one photon
pair. Furthermore, this process is triggered by applying a potential
difference over the microcavity, which allows for greater control over
the creation of a pair. In particular, the probability of creating a
pair can be as high as $p_{\rm s} = 0.9$. Consequently, this source
satisfies the required double-photon gun properties outlined above. 

The two entangled photons from this source have different frequencies,
which allows us to spatially separate them by means of a dichroic
mirror. Interference phenomena at beam splitters, however, rely on the
indistinguishability of the incoming photons, and the non-degenerate
frequencies might render the photons distinguishable. Special care
needs to be taken to arrange the setup in such a way that only photons
with equal frequencies enter any particular optical element.

In practice, quantum communication protocols may use any of the four
two-qubit Bell states: $|\Phi^{\pm}\rangle = (|H,H\rangle \pm
|V,V\rangle)/\sqrt{2}$ and $|\Psi^{\pm}\rangle = (|H,V\rangle \pm
|V,H\rangle)/\sqrt{2}$, with $H$ and $V$ the polarization directions.
These states are locally transformed into each other by means of
simple qubit operations, and we therefore may assume that the above
entanglement source (entangler) can make any of the four Bell
states. Now let us look at the other ingredients of our quantum
repeater.  


The entanglement-swapping component (swapper) of the quantum repeater
is essentially nothing more than a Bell detector. It is well known
that it is impossible to make a deterministic, complete, Bell
measurement with linear optics \cite{lutkenhaus99}, but one can
distinguish two out of four two-qubit Bell states with a simple beam
splitter configuration \cite{braunstein95}. Recently, Franson and
co-workers have shown that a controlled-not (CNOT) ---and hence a Bell
measurement--- is possible {\em probabilistically} with only
projective measurements and entangled input states
\cite{pittman01}. The probability of success for this CNOT is not
large enough to make the Bell measurement more efficient, but it will
be an essential component of our purifier.  

\begin{figure}[h]
  \begin{center}
  \begin{psfrags}
     \psfrag{D1}{$D_1$}
     \psfrag{D2}{$D_2$}
     \psfrag{phi}{$\!\!|\Phi^+\rangle$}
     \epsfxsize=8in
     \epsfbox[-100 20 1200 185]{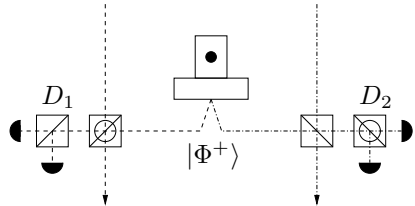}
  \end{psfrags}
  \end{center}
  \caption{The probabilistic CNOT gate. Conditioned on a specific
     detector outcome in $D_1$ and $D_2$, the setup performs a
     controlled not. The boxed beam splitter is a linear polarization
     beam splitter and the circled box is a circular polarization beam
     splitter.}  
  \label{fig2}
\end{figure}


In order to distill maximally entangled quantum states, one needs
entanglement purification. Suppose Alice and Bob share two pairs of
non-maximally entangled states. Bennett {\em et al}.\ showed that with 
some finite probability it is possible to extract a single maximally
entangled state \cite{bennett96}. To do this, both Alice and Bob apply 
a CNOT, where the halves of the first entangled state pair serve as
the control qubit, and the halves of the second as the target. The
target qubits are then measured in the computational basis (determined
by the participants prior to execution, e.g., $|H\rangle$ and
$|V\rangle$), and conditioned on a parallel coincidence (like
$|H\rangle_A |H\rangle_B$ and $|V\rangle_A |V\rangle_B$), Alice and
Bob now share a maximally entangled state in the remaining two
qubits. The probability of purification depends on the fidelity of the
incoming entangled state, and therefore on the channel noise factor
$\gamma$. 

Additionally, in some cases the modes of the control qubit might be
empty, because the entanglement source failed to create a photon. In
order to rule out these events, we can employ the single-photon
quantum nondemolition ({\sc sp-QND}) measurement scheme proposed by
Kok {\em et al}.\ \cite{kok02}. This is a probabilistic scheme that
can be set up to signal a single photon in an optical mode without
destroying its polarization. The success rate of this device is
$p_{\mathsc qnd} = \frac{1}{8}$. This device employs four
photodetectors, as well as two entanglement sources to create the
auxiliary single-photon input states. 


Essential for the success of this protocol is the ability to perform
the controlled-not operation. In Fig.\ \ref{fig2} we show the
schematic setup for the probabilistic CNOT designed by Pittman {\em et
  al}.\ \cite{pittman01}. The main ingredients are a $|\Phi^+\rangle$
source and four polarization beam splitters, two of which separate
{\em circular} polarization. The control qubit enters a linear
polarizing beam splitter, and the target enters a circular polarizing
beam splitter. The secondary input ports of these two beam splitters
are fed by the two components of a $|\Phi^+\rangle$ Bell state.  

A successful CNOT operation is now conditioned on detecting a linearly 
polarized photon after the circularly polarizing beam splitter ($D_1$ 
in Fig.\ \ref{fig2}), and a circularly polarized photon after the
linearly polarizing beam splitter ($D_2$ in Fig.\ \ref{fig2}). These 
detections can be implemented with suitable polarizing beam splitters
and ordinary photodetectors \cite{pittman01}. The probability of this
CNOT operation is given by $p_{\mathsc cnot}=\frac{1}{4}$. 

\end{multicols}
\noindent\rule{5cm}{0.5pt}
\begin{figure}[h]
  \begin{center}
  \begin{psfrags}
     \psfrag{E}{\sf E}
     \psfrag{P}{\sf P}
     \psfrag{S}{\sf S}
     \psfrag{q}{\small\sc qnd}
     \psfrag{is}{$=$}
     \psfrag{.}{}
     \epsfxsize=8in
     \epsfbox[-100 0 700 150]{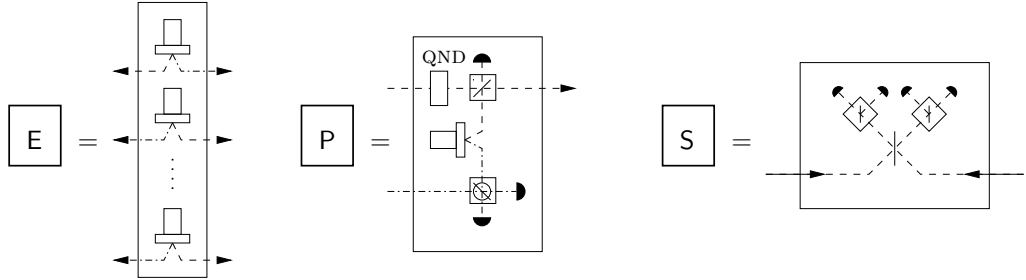}
  \end{psfrags}
  \end{center}
  \caption{The components of the quantum repeater. The three boxes
     {\sf E}, {\sf P}, and {\sf S} denote the entanglement station
     (entangler), the purifier and the swapping element (swapper)
     respectively. The entanglers are drawn as little top hats. The
     dashed and the dash-dotted lines represent the fact that the two
     output modes have different frequencies. The purifier element
     contains a QND device, an optical CNOT gate, and a detector on
     one output mode. The swapper implements a partial Bell
     measurement.}   
  \label{fig3}
\end{figure}


\begin{figure}[h]
  \begin{center}
  \begin{psfrags}
     \psfrag{E}{$\!$\sf E}
     \psfrag{P}{$\!$\sf P}
     \psfrag{S}{\sf S}
     \psfrag{Alice}{Alice}
     \psfrag{Bob}{Bob}
     \epsfxsize=8in
     \epsfbox[-30 0 770 75]{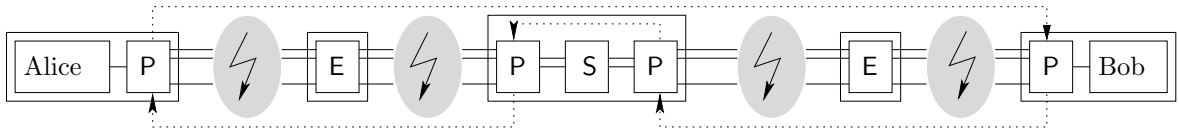}
  \end{psfrags}
  \end{center}
  \caption{The assembled quantum repeater. The solid lines represent
     the quantum channels and the dotted lines denote classical
     communication channels. The distance is included in the shaded
     region. This is also where the ``decoherence devil'' resides [19].
     Note that the entanglement sources are separate
     stations. Alternatively, the entanglement station can be placed
     near Alice and Bob.} 
  \label{fig4}
\end{figure}
\hfill\rule{5cm}{0.5pt}
\begin{multicols}{2}   

In order to build a complete quantum repeater, we have to integrate
the components described above into a circuit \cite{briegel98,dur99}. 
The separate components are (see Fig.\ \ref{fig3}) the {\em entangler}
{\sf E}, the {\em purifier} {\sf P} and the {\em swapper} {\sf S}.  
The assembled quantum repeater, shown in Fig.\ \ref{fig4}, is a circuit
involving {\sf E}, {\sf P} and {\sf S}, together with classical
communication between the different stations. This classical channel
is necessary to exchange information about the measurement outcomes of
the purifiers and about the location of the purified (and swapped)
entanglement. 


Finally, we address the success rate of the quantum repeater. The
success rate is given by the reciprocal of the single-pair creation
rate between Alice and Bob. It turns out that this value is highly 
dependent on the quantum efficiency of the photodetectors. 

To purify two entangled photon pairs, we have to take into account the
probability of success for the individual components, as well as the
losses in the system. Suppose we have six double-photon guns (two for
the photon sources, two in the QND device and one for every CNOT) and
ten detectors with quantum efficiency $\eta$ (three per CNOT and four
in the QND device). Furthermore, let the noise parameters due to the
attenuation be given by $\gamma$ for the dephasing (reducing the
fidelity) and $\zeta$ for the photon loss over the channel. The
probability for purifying a single pair of entangled photons is then
given by 
\begin{equation}\label{purrate}
  p_{\rm pur} = p_{\rm s}^6\, \eta^{10}\, (1-\gamma)^2\,
  \zeta^2\, p_{\mathsc cnot}^2\, p_{\mathsc qnd}\; ,
\end{equation}
where $p_{\rm s}$ is the probability of success of the double-photon
gun. It is immediately obvious that a reduced quantum efficiency
$\eta$ will strongly contribute to the deterioration of the success
rate, du to the $\eta^{10}$ behaviour.  

In order to make a repeater, we start with two purified pairs and
perform entanglement swapping on their two halves. This swapping
protocol is not deterministic, and is subject to losses as well. As can
be seen in Fig.\ \ref{fig3}, the swapping element requires a two-fold
detector coincidence. Furthermore, a complete Bell detection occurs
only 50\% of the time. The probability of success for entanglement
swapping is therefore given by 
\begin{equation}
  p_{\rm swap} = \frac{\eta^2}{2}\; .
\end{equation}

Let us now insert some values of the several components. We will use
different values for the detector efficiency, since this is the most
important parameter. Choose for example $p_{\rm s}=0.9$,
$\gamma=\frac{1}{2}$, $\zeta=\frac{1}{2}\sqrt{2}$, $p_{\mathsc
  cnot}=\frac{1}{4}$ and $p_{\mathsc qnd}=\frac{1}{8}$. For three
different values of $\eta$, this gives rise to Table \ref{tab1} (with
$N_{\rm pur} = p_{\rm pur}^{-1}$ and $N_{\rm swap} = p_{\rm
  swap}^{-1}$). Since a repeater needs two purifiers and one swapper,
the total number of components $N_{\rm total}$ is given by $N_{\rm
  total} = 2N_{\rm pur}N_{\rm swap}$. The results of Table \ref{tab1}
should be compared with the number of transistors on a Pentium chip,
which is of the order $10^7$. We might reduce the number of components
if we use the entanglers and purifiers {\em in series}. However, this
would require a quantum memory or a classical delay line for the
entangled photons. 

It is immediately clear that an improvement in the detector efficiency
yields a substantial gain in the efficiency of the protocol, due to
the factor $\eta^{10}$ in Eq.\ \ref{purrate}. Even though detector
efficiencies of 0.8 are quoted, experimental values are as bad as
0.3. Therefore, in order to operate the repeater more efficiently,
better detectors are needed.

Note also that intelligent switching, conditioned on detector outcomes
and classical communication between the components, is needed both to
purify and to correlate purified entanglement in the swapping
procedure. This results in an overhead in the number of components.

In conclusion, we proposed a practical implementation for a quantum
repeater employing double-photon guns, probabilistic CNOT operations,
and quantum nondemolition measurements. The protocol utilizes
available and almost available technology. Possible drawbacks are the
conditional switching and the low quantum efficiencies of
state-of-the-art photodetectors. 

This work was carried out at the Jet Propulsion Laboratory, California
Institute of Technology, under a contract with the National Aeronautics 
and Space Administration. The authors acknowledge J.D.\ Franson, H.\
Lee, G.M.\ Hockney and D.V.\ Strekalov for valuable discussions. In
addition, P.K.\ acknowledges the United States National Research
Council. Support was received from the Office of Naval Research, 
Advanced Research and Development Activity, National Security Agency,
and the Defense Advanced Research Projects Agency.  

\begin{table}[h]
\begin{center}
\begin{tabular}{l||l|l|l}
 $~\eta$ & $N_{\rm pur}~$ & $N_{\rm swap~}$ & $N_{\rm total}~$ \cr\hline
 ~0.3 & $3\cdot 10^{7}$ & 20 & $\sim 10^{9} \phantom{\rule{0pt}{12pt}}$ \cr
 ~0.8 & $2\cdot 10^{3}$ & 3 & $\sim 10^{4}$ \cr
 ~1 & 250 & 2 & $\sim 10^{3}$ 
\end{tabular}
\end{center}
\caption{The number of components in the quantum repeater for
 different values of the detector efficiency $\eta$.}
\label{tab1}
\end{table}

\end{multicols}
\end{document}